\documentclass[aps,pre,reprint,longbibliography,floatfix,superscriptaddress]{revtex4-1}
\usepackage{graphicx}
\usepackage{amsmath,amssymb,amsfonts}
\usepackage{hyperref}

\begin{document}

\title[Paris car parking problem]{``Paris car parking problem'' for partially ordered discorectangles on a line}

\author{Nikolai I. Lebovka}
\email[Corresponding author: ]{lebovka@gmail.com}
\affiliation{Department of Physical Chemistry of Disperse Minerals, F. D. Ovcharenko Institute of Biocolloidal Chemistry, NAS of Ukraine, Kyiv 03142, Ukraine}
\affiliation{Department of Physics, Taras Shevchenko Kiev National University, Kyiv 01033, Ukraine}

\author{Mykhailo O. Tatochenko}
\affiliation{Mykolaiv Professional Shipbuilding Lyceum, Mykolaiv 54011, Ukraine}

\author{Nikolai V. Vygornitskii}
\affiliation{Department of Physical Chemistry of Disperse Minerals, F. D. Ovcharenko Institute of Biocolloidal Chemistry, NAS of Ukraine, Kyiv 03142, Ukraine}

\author{Yuri Yu. Tarasevich}
\email[Corresponding author: ]{tarasevich@asu.edu.ru}
\affiliation{Laboratory of Mathematical Modeling, Astrakhan State University, Astrakhan 414056, Russia}

\date{\today}

\begin{abstract}
The random sequential adsorption (RSA) of identical elongated particles (discorectangles) on a line (``Paris car parking problem'') was studied numerically. An off-lattice model with continuous positional and orientational degrees of freedom was considered.
The possible orientations of the discorectanles were restricted between $\theta \in [-\theta_\text{m};\theta_\text{m}]$ while the aspect ratio (length-to-width ratio) for the discorectangles was varied within the range $\varepsilon \in [1;100]$. Additionally, the limiting case $\varepsilon=\infty$ (i.e., widthless sticks) was considered. We observed, that the RSA deposition for the problem under consideration was governed by the formation of rarefied holes (containing particles oriented along a line) surrounded by comparatively dense stacks (filled with almost parallel particles oriented in the vertical direction). The kinetics of the changes of the order parameter, and the packing density are discussed. Partial ordering of the discorectangles significantly affected the packing density at the jamming state, $\varphi_\text{j}$, and shifted the cusps in the $\varphi_\text{j}(\varepsilon)$ dependencies. This can be explained by the effects on the competition between the particles' orientational degrees of freedom and the excluded volume effects.
\end{abstract}

\maketitle

\section{Introduction\label{sec:intro}}

The random sequential adsorption model (RSA) is widely used for modeling the packing of particles in spaces of different dimensionalities~\cite{Evans1993,Torquato2010}. In this model, particles are placed randomly and sequentially. Their overlapping with previously placed particles is strictly forbidden, i.e., excluded volume interaction between particles is assumed. The basic variant of the RSA model also assumes the absence of any relaxation, diffusion and desorption. Sequential placing of particles leads to the formation of a jammed state where no additional particle can be added due to the absence of appropriate holes. Different variants of one-dimensional (1D) and higher dimensional [e.g., two-dimensional (2D)] models have previously been numerically studied~\cite{Cadilhe2007,Privman2016}. The 2D RSA model has been particularly widely used for simulation of the adsorption of colloids and proteins~\cite{Adamczyk2012}.

The 1D self-assembly continuously attracts growing theoretical interest and technological demand. The attractive properties of low-dimensional functional materials were intensively studied in different fields of modern material science and nanotechnology. The 1D composite nanomaterials can be used in electronic and optical nanodevices, chemical and biological sensors, and environment, energy, and biomedical fields~\cite{Lu2009Small,Devan2012AFM,Zhai2012}. The different strategies were applied to prepare the 1D crystalline nanoarrays~\cite{Lu2004JACS}. The 1D assembly of metal nanoparticles (so-called nanochains) were fabricated for tunable surface plasmon resonance properties~\cite{Yang2006}. Insertion of organic dyes into the 1D channels of zeolite L allowed producing the functional composites with intriguing optoelectronic and photochemical properties~\cite{Fois2012JPhChC,Tabacchi2016ChemComm,SolaLlano2019}. The properties of such composites significantly depend on orientations and arrangements of particles in 1D channels~\cite{Fois2012JPhChC}. The incorporation of bioorganic systems (protein-containing water nanodroplets) in a porous inside 1D silica nanochannels was studied~\cite{Giussani2019}. The 1D confined supramolecular architectures of chromophores can be used for solar energy harvesting and storage~\cite{Fois2019}.

For 1D RSA packing onto a line (the so-called car parking problem), an analytical description of the processes can be obtained in many cases~\cite{Talbot2000,Krapivsky2010}. For example, the kinetics of the RSA deposition of equal disks (continuum RSA problem) is described by the following equation~\cite{Renyi1963,Gonzalez1974}
\begin{equation}\label{eq:Renyi}
  \varphi(t) =\int_{0}^{t} \exp\left[-2\int_{0}^{x} y^{-1} (1 - \mathrm{e}^{-y})\,\mathrm{d}y\right]\,\mathrm{d}x,
\end{equation}
where $\varphi(t)$ is the packing density (coverage). When approaching the jamming limit ($t \to \infty$), the terminal packing coverage $\varphi (t)$ demonstrates the algebraic time dependence~\cite{Pomeau1980,Krapivsky1992}:
\begin{equation}\label{eq:lambda}
\varphi (t)=\varphi_\text{j}-\exp(-2\gamma )t^{-\nu},
\end{equation}
where $\gamma=0.577215\dots $ is Euler's constant and $\varphi_\text{j}=C_\text{R}=0.7476\dots$ is the famous R\'{e}nyi's parking constant~\cite{Renyi1963}. $\nu = 1/d_\text{f}$, where $d_\text{f}$ is related to the number of degree of freedom for a deposited object. $d_\text{f}$ coincides with the dimensionality of the system when isotropic particles are deposited in continuous media.

An analytical expression of the density pair distribution function $g_2(r)$ for 1D RSA packing has also been derived~\cite{Bonnier1994}, and the result compared with the well-known Frenkel's result for the continuum equilibrium  problem of 1D fluid~\cite{Frenkel1946,Salsburg1953}.

At jamming concentration (i.e., at $\varphi=\varphi_\text{j}=C_\text{R}$), a comparison of the pair distribution functions derived for the RSA [$g_2(r)$] and equilibrium fluid [$g_2^e(r)$] problems revealed the influence of the irreversibility on the near-neighbor correlations~\cite{Bonnier1994}. The equilibrium function $g_2^e(r)$ displayed a relatively large correlated region (large oscillations), whereas the RSA function $g_2(r)$ was short-ranged  and close to unity at $r>3$.

The 1D RSA problems of randomly oriented particles with arbitrary shapes (ellipses, rectangles, discorectangles, etc.) have also been analyzed~\cite{Chaikin2006,Baule2017,Ciesla2020}. This case is commonly referred to as the ``Paris car parking problem''. With increase of the aspect ratio $\varepsilon$ (the length-to-diameter ratio) the jamming (maximum) coverage  increased from the value $\varphi_\text{j}= 0.7476$ for circles ($\varepsilon=1)$, went through a maximum (cusp) at some value of $\varepsilon_\text{m} \approx 1.5$, and then decreased at  higher aspect ratios~\cite{Chaikin2006}. Such behavior has been observed in the packing of randomly oriented ellipses~\cite{Chaikin2006} and later on for ellipses ($\varepsilon_\text{m} \approx 1.5$, $\varphi_\text{j}=0.775380 \pm 0.000019$), rectangles ($\varepsilon_\text{m} \approx1.3$, $\varphi_\text{j}=0.749575 \pm 0.000016$), and discorectangles ($\varepsilon_\text{m} \approx1.5$, $\varphi_\text{j}=0.781249 \pm 0.000020$)~\cite{Ciesla2020}.
The initial coverage increase was explained by the relaxing of a parameter constraint (orientation of particles), while the coverage decrease above $\varepsilon_\text{m}$ was explained by excluded volume effects~\cite{Chaikin2006}. Universality in the jamming limit for elongated particles (e.g., needles, rectangles, ellipses) in 1D systems has also been discussed~\cite{ Kantor2009}. In particular, the dependence of the universality class on the object's shape was demonstrated.

Similar $\varphi_\text{j}(\varepsilon)$  behavior was also observed in the RSA packing of elongated particles in 2D and 3D systems~\cite{Chaikin2006}.
For the ``Paris car parking problem'', when approaching the jamming limit ($t \to \infty$), an  algebraic time dependence $\varphi(t) \propto t^{-\nu}$ was observed~\cite{Baule2017}. The exponent $\nu = 1/d_\text{f}$ was dependent on the particle shape and the number of orientational degrees of freedom of each particle. For ellipses, the simulated empirical exponent was $d_\text{f} = 1.5$, while, for discorectangles and rectangles, they were in the range $1.5 \leqslant d_\text{f} \leqslant 2$. More recently, it has been demonstrated that $d_\text{f} = 1.5$ and $2.0$, respectively, for packings built of ellipses and rectangles~\cite{Ciesla2020}. For discorectangles of moderate aspect ratio, $\varepsilon$, a continuous transition between these two values was observed, from $d_\text{f} = 1.5$ at relatively small $\varepsilon$ values to $d_\text{f} = 2$ at large $\varepsilon$ values ($\varepsilon \gg 1$).

However, all previous studies of the ``Paris car parking problem'' have been devoted to conventional RSA with random orientation of elongated particles and have not paid attention to the effects of the particle orientation constraint on the kinetics and structure of the deposits.

This paper analyzes numerically RSA of identical elongated particles (discorectangles) on a line (``Paris car parking problem'') using an off-lattice model with continuous positional and orientational degrees of freedom. Special attention has been paid to the effect of the constraint of particle orientation on the packing. The kinetics of changes in the order parameter and the packing density are discussed.  The rest of the paper is constructed as follows. In Sec.~\ref{sec:methods}, the technical details of the simulations are described, all necessary quantities are defined, and some test results are given. Section~\ref{sec:results} presents our principal findings. Section~\ref{sec:conclusion} summarizes the main results.

\section{Computational model\label{sec:methods}}
RSA packing of discorectangles were generated using a saturated packing algorithm similar to that developed for jammed packings of non-oriented anisotropic objects on a 2D plane~\cite{Haiduk2018} or a 1D line~\cite{Ciesla2020}. It is based on tracing regions where subsequent particles can be added. The centers of the particles were placed on the 1D line and periodic boundary conditions were used to minimize any finite-size effects.

\begin{figure}[!htbp]
	\centering
	\includegraphics[width=\columnwidth]{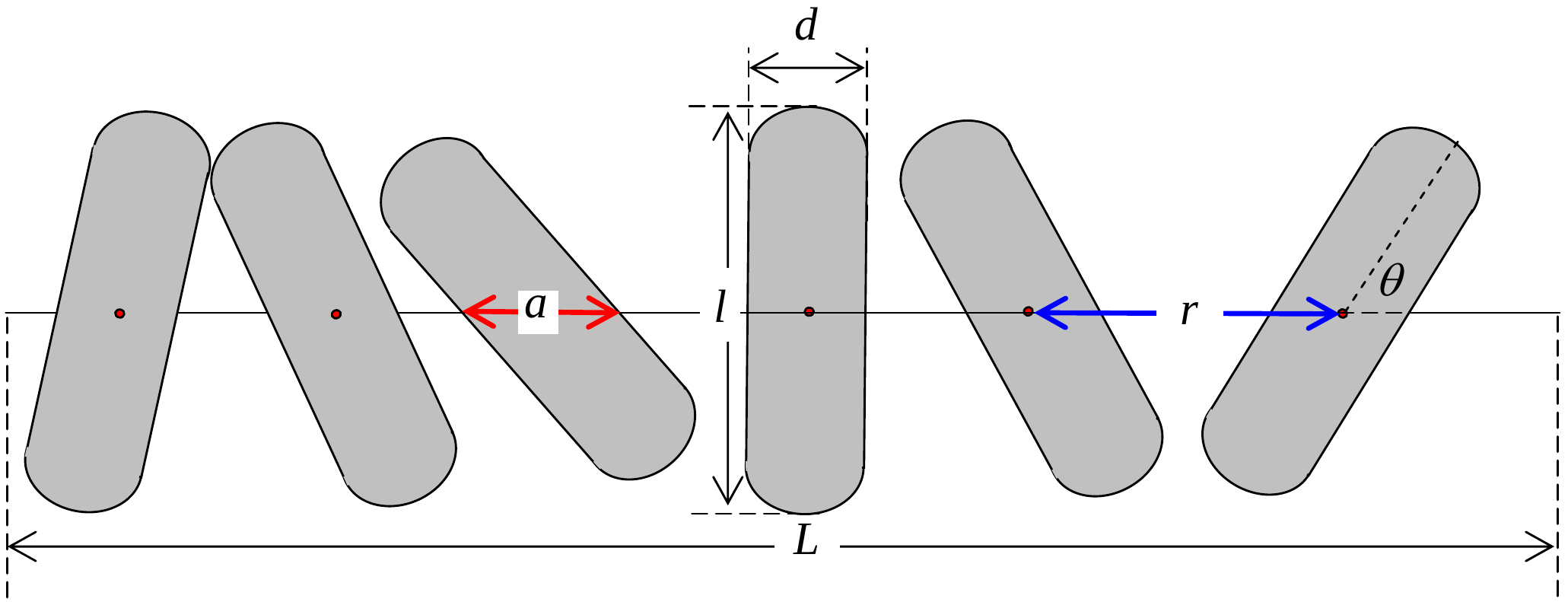}
	\caption{A description of the RSA packing of discorectangles on a line. Intersections of the particles are forbidden. Each deposited particle covers a distance $a$ on the line. Here $L$ is the total length of the line, and $r$ is the distance between the particle centers.\label{fig:f01}}
\end{figure}

The aspect ratio (length-to-width ratio) of discorectangles was defined as $\varepsilon=l/d$ (Fig.~\ref{fig:f01}). Both infinitely thin particles (sticks) with $ \varepsilon = \infty$ and discorectangles with $\varepsilon \in [1;100]$ were considered.

The orientation of the particles was characterized by the mean order parameter defined as
\begin{equation}\label{eq:S}
  S = \left\langle \cos 2\theta  \right\rangle,
\end{equation}
where $\langle\cdot\rangle$ denotes the average, $\theta$ is the angle between the long axis of the particle and the director $n$, which gives the direction of the preferred orientation of the particles. Note, that $S=1$ and $S=-1$ correspond to ideally oriented particles along the line and perpendicular to it, respectively.

To simulate the general case, a model of anisotropic random-orientation distribution was used~\cite{Balberg1983}. For this model, the orientations of the deposited particles are selected to be uniformly distributed within some interval such that $-\theta_\text{m} \leqslant \theta \leqslant \theta_\text{m}$, where $\theta_\text{m} \leqslant \pi/2$. For this model, the preassigned order parameter can be evaluated as~\cite{Lebovka2019}
\begin{equation}\label{eq:S0}
S_0 = \frac{\sin 2\theta_\text{m}}{2\theta_\text{m}},
\end{equation}
and for the isotropic case ($\theta_\text{m} = \pi/2$), it is given by $S_0=0$. During the deposition, some particle orientations may be rejected and the real order parameter in the deposit, $S$, differs from the value of $S_0$. The situation is similar to that observed in the RSA deposition of partially oriented elongated particles ($k$-mers) on square lattice~\cite{Lebovka2011}.

All distances were measured in units of particle length, while time was measured using dimensionless time units, $t=n/L$, where $n$ is the number of deposition attempts, and $L$ is the total length of the line. Each deposited particle covers a distance $a$ on the line, thus, the average coverage (packing density) was defined as $\varphi = L^{-1}\sum_i a_i$, where the summation goes over all particles.

The density and orientational pair correlations were characterized using the distribution functions
\begin{align}\label{eq:g2s2}
  g_2(r) & =C_2(r)/C_2(\infty),\\
  s_2(r) & =S_2(r)/S_2(\infty),
\end{align}
where
\begin{align}\label{eq:C2S2}
  C_2(r) &= \langle \rho(0)\rho(r)\rangle, \\
  S_2(r) &= \langle \cos\{2[\theta(0)-\theta(r)]\}\rangle,	
\end{align}
$\rho(r)$  is the local number density and $r$ is the distance between the particle centers (Fig.~\ref{fig:f01}). The non-zero asymptotic value of $S_2(\infty)$ suggests a long-range nematic order.

The average number density was calculated as $\rho_0=N/L$, where $N$ is the total number of deposited particles. To determine the effects of system size, finite size scaling analysis for $L$ in the interval $L \in [2^{12};2^{15}]$ ($L \in [4096;32768]$) was performed. Typically, the length of the line was taken as $L = 2^{15}=32768$. The jamming coverage was assumed to be reached after at least $L\times 10^{10}$ unsuccessful attempts to place a new particle on the line. For each given value of $\varepsilon$ and $S_0$, the computer experiments were repeated using from 10 to 20 independent runs. The error bars in the figures correspond to the standard deviation of the mean. When not shown explicitly, they are of the order of the marker size.

\section{Results and Discussion\label{sec:results}}
\subsection{Sticks}

\begin{figure}[!htbp]
	\centering
	\includegraphics[width=\columnwidth]{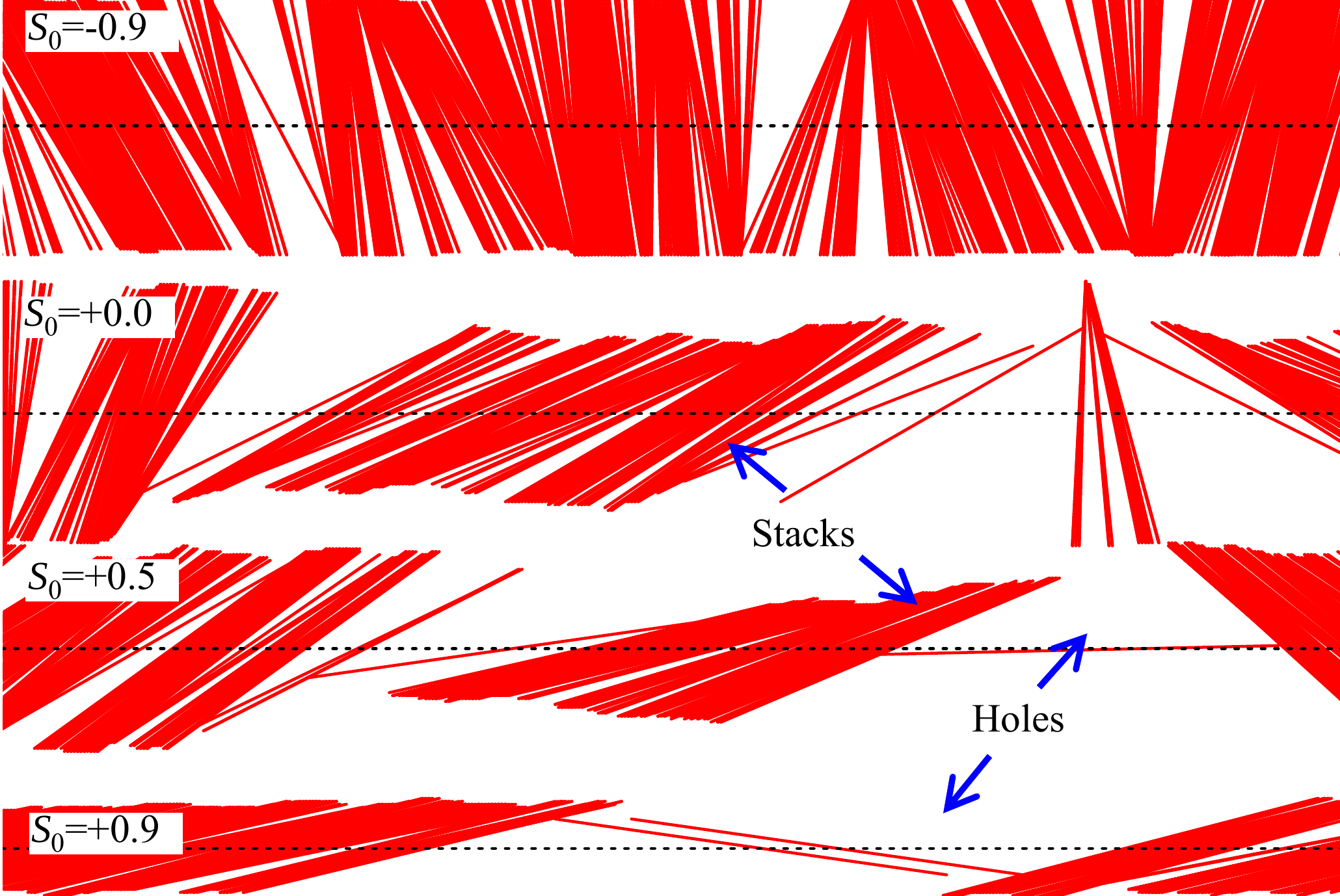}
	\caption{Examples of packing patterns of infinitely thin particles (sticks) with $\varepsilon = \infty$ at different values of the preassigned order parameter $S_0$. Length of the line is $L=2^{15}$ and time is $t=10^{10}$. \label{fig:f02}}
\end{figure}

Figure~\ref{fig:f02} shows examples of the packing patterns of infinitely thin particles (sticks) with $\varepsilon = \infty$ at different values of the preassigned order parameter $S_0$.  At negative values of $S_0$, the formation of dense stacks of sticks oriented in the vertical direction was observed. Even at relatively large orientational ordering (e.g., at $S_0=-0.9$), small holes between stacks were observed. At $S_0 \geqslant 0$, the holes could occupy a significant fraction of the line space.

\begin{figure}[!htbp]
	\centering	
\includegraphics[width=\columnwidth]{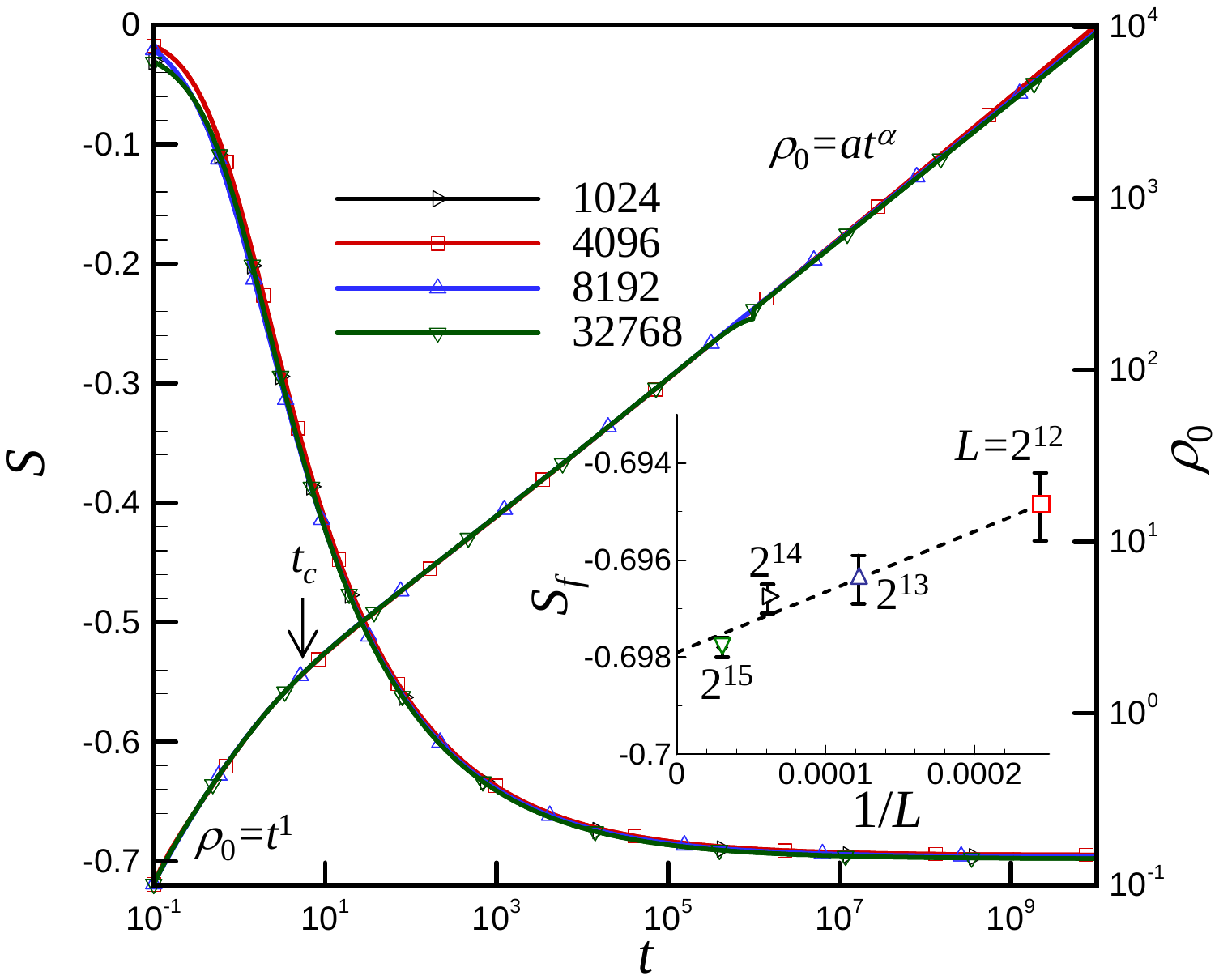}\\
	\caption{The order parameter, $S$, and mean number density, $\rho_0$, versus the deposition time, $t$, at a zero value of the preassigned order parameter, $S_0=0$. The data are presented for deposition of sticks ($\varepsilon = \infty$) on the  line with length $L$.  Inset  presents the limiting order parameter $S_\text{f}$ (at $t \to \infty$) versus the inverse length of the line $1/L$.  \label{fig:f03}}
\end{figure}

Figure~\ref{fig:f03} demonstrates the kinetics of the order parameter, $S$, and mean number density, $\rho_0$, at a zero value of the preassigned order parameter, $S_0=0$, and different values of the line length, $L$. The conventional RSA model does not allow preservation of the preassigned order parameter $S_0$. In this model, the line substrate with previously deposited particles ``selects'' a newcomer stick with appropriate orientation, so this would result in a  deviation of the preassigned order parameter, $S_0$, from the actually obtained one, $S$. The situation is rather similar to that observed for RSA deposition of partially oriented elongated particles on a square lattice~\cite{Lebovka2011}. Analysis of the deposition patterns presented in Fig.~\ref{fig:f02} evidences that the particles deposited in an almost horizontal direction (inclined along the line) form holes and block the further deposition along the line. However, particles deposited in the almost vertical directions promote further deposition inside stacks and serve as attractors for almost vertical deposition. Therefore, the line substrate can serve as a filter for particles with appropriated orientation.

The actual value of $S$ gradually decreased with increasing time, $t$, approaching the value $S_\text{f}$ in the limit of an infinitely large time, $t \to\infty$. This reflects the formation of stacks of particles arranged perpendicularly to the deposition line (Fig.~\ref{fig:f02}, $S_0=0$).
Analysis of scaling in the coordinates $S_\text{f}$ versus $1/L$ allows estimation of the value of the order parameter for infinitely large system, $S_\text{f} = -0.697 \pm 0.009$ (Fig.~\ref{fig:f03}). This value is only slightly different from the value $S_\text{f} = -0.6977 \pm 0.002$ obtained at $L=2^{15}$. The similar behavior was also observed for finite values of $\varepsilon$. For this reason, in this work, the studies were mainly performed at $L=2^{15}$  without additional scaling analysis.

\begin{figure}[!htbp]
	\centering	
\includegraphics[width=\columnwidth]{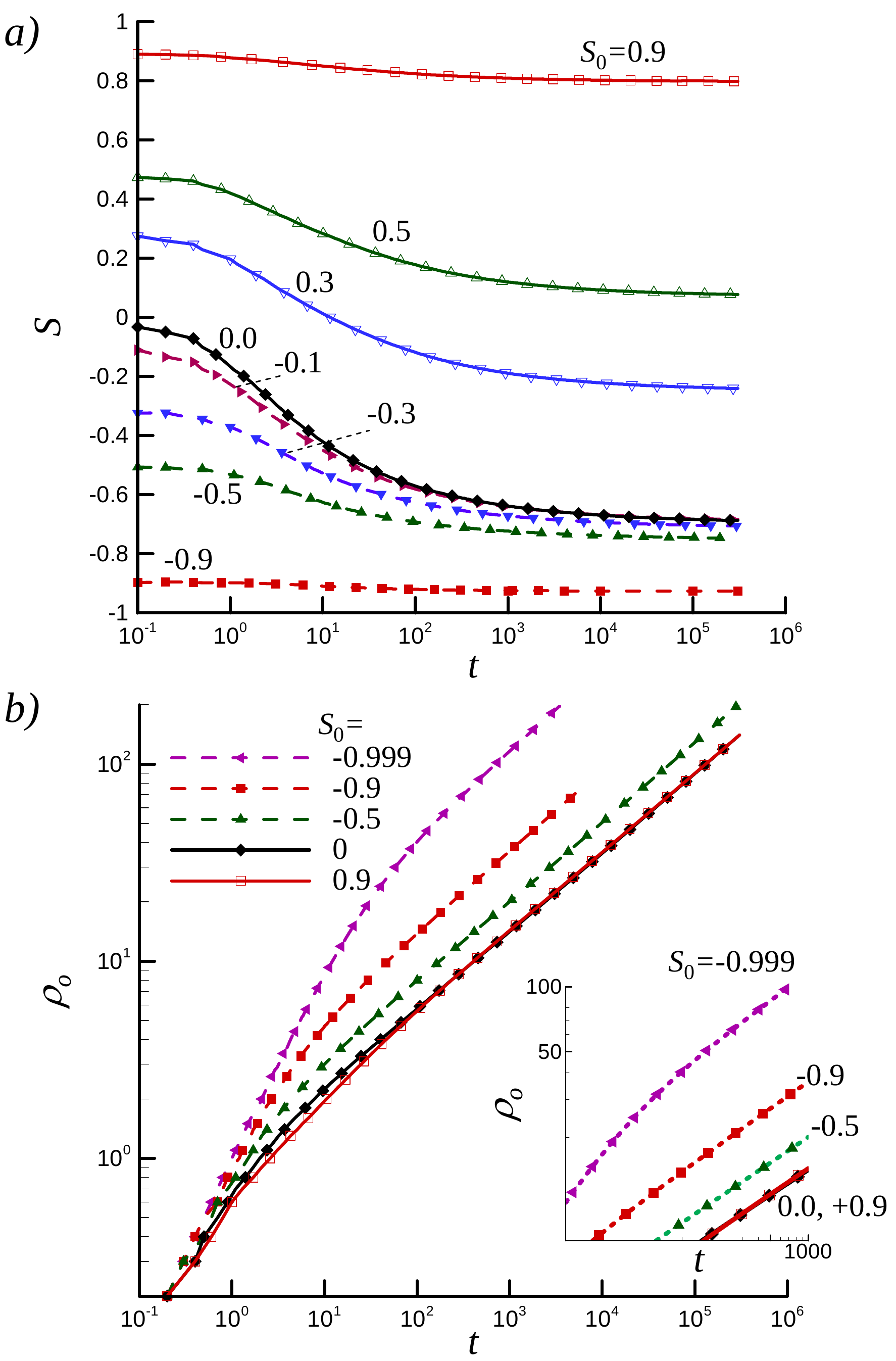}\\
	\caption{Order parameter, $S$, (a) and mean number density, $\rho_0$, (b) versus the deposition time, $t$, at different preassigned order parameters, $S_0$. The data are presented for the deposition of sticks ($\varepsilon = \infty$) on a line with length $L=2^{15}$. Inset in (b) presents the enlarged portion of the $\rho_0(t)$ dependencies.\label{fig:f04}}
\end{figure}

The mean number density, $\rho_0$, continuously increases with packing time, $t$. The time dependence $\rho_0(t)$ demonstrated clear crossover behavior. For loose packings at initial times, below $t_\text{c} \approx 1$, almost linear dependencies $\rho_0 \approx t$ were observed. However, at $t>t_\text{c}$, the data can be well fitted by the power law dependence $\rho_0 = A t^\alpha$, where $A = 0.8813 \pm 0.0002$ and $\alpha =0.40022 \pm 0.0003 \approx 2/5$ are the fitting parameters ($R^2=0.99994$). From these data the transition time can be estimated as $t_\text{c}=A^{1-\alpha} \approx 0.93$. Figure~\ref{fig:f04} shows the order parameter, $S$, (a) and mean number density, $\rho_0$, (b) versus the deposition time, $t$, at different preassigned order parameters, $S_0$. The data are presented for sticks ($\varepsilon = \infty$) and the line length is $L=2^{15}$. The inset in Fig.~\ref{fig:f04}b presents an enlarged portion of the $\rho_0(t)$ dependencies. Across the studied range of $S_0$ ($-1 \leqslant S_0 \leqslant 1$), the values of $S$ gradually decreased with increasing time, $t$, and, finally, in the limit of an infinitely large time, they reached values of $S_\text{f}$ smaller than the preassigned values, $S_0$ (Fig.~\ref{fig:f04}a). This significant dropping of order parameters began after the formation of stacks could be observed at intermediate deposition times. It can be explained by a manifestation of the filtering properties of line substrates.  The crossover behavior in $\rho_0(t)$ was observed at different values of the preassigned order parameters, $S_0$ (Fig.~\ref{fig:f04}b). At long deposition times, the scaling exponents were approximately the same for different values of $S_0$ ($S_0$ ($-1 \leqslant S_0 \leqslant 1$), $\alpha \approx 2/5$. However, the transition time $t_\text{c}$ was dependent upon the value of  $S_0$ and $t_\text{c} \to \infty$ in the limit of $S_0 \to -1$.

\begin{figure}[!htbp]
	\centering	
\includegraphics[width=\columnwidth]{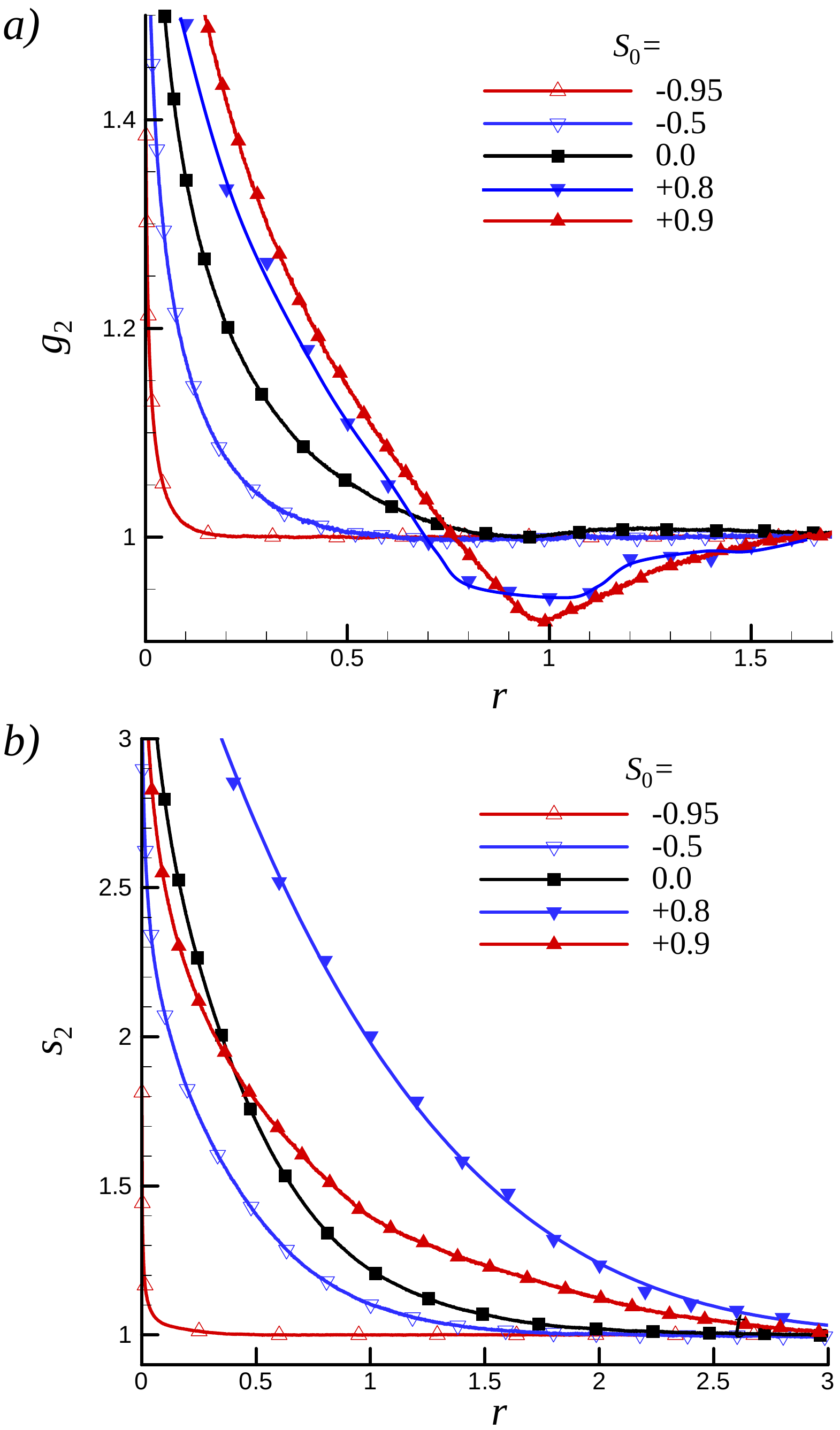}\\
	\caption{The density $g_2(r)$ (a) and orientation $s_2(r)$ (b) pair correlation functions for different values of the preassigned order parameter, $S_0$. The data are presented for the deposition of sticks ($\varepsilon = \infty$) on a line with length $L=2^{15}$, and a deposition time $t=10^6$. \label{fig:f05}}
\end{figure}

Figure~\ref{fig:f05} presents examples of the density $g_2(r)$ (a) and orientation $s_2(r)$ (b) pair correlation functions for different values of the preassigned order parameter, $S_0$. Both  the functions $g_2(r)$ and $s_2(r)$ exhibit a strong divergence in the limit of $r \to 0$. This corresponds to the formation of stacks with parallel arrangements of sticks (Fig.~\ref{fig:f01}). Moreover, at $S_0 \geqslant 0$, the density pair correlation functions $g_2(r)$ demonstrate oscillations that asymptotically approach $g(r)=1$ at large distances. The pronounced minimums observed in the range of $r \in [0.5;1.1]$ corresponds to the correlations between places filled with stacks and holes.

\begin{figure}[!htbp]
	\centering
	\includegraphics[width=\columnwidth]{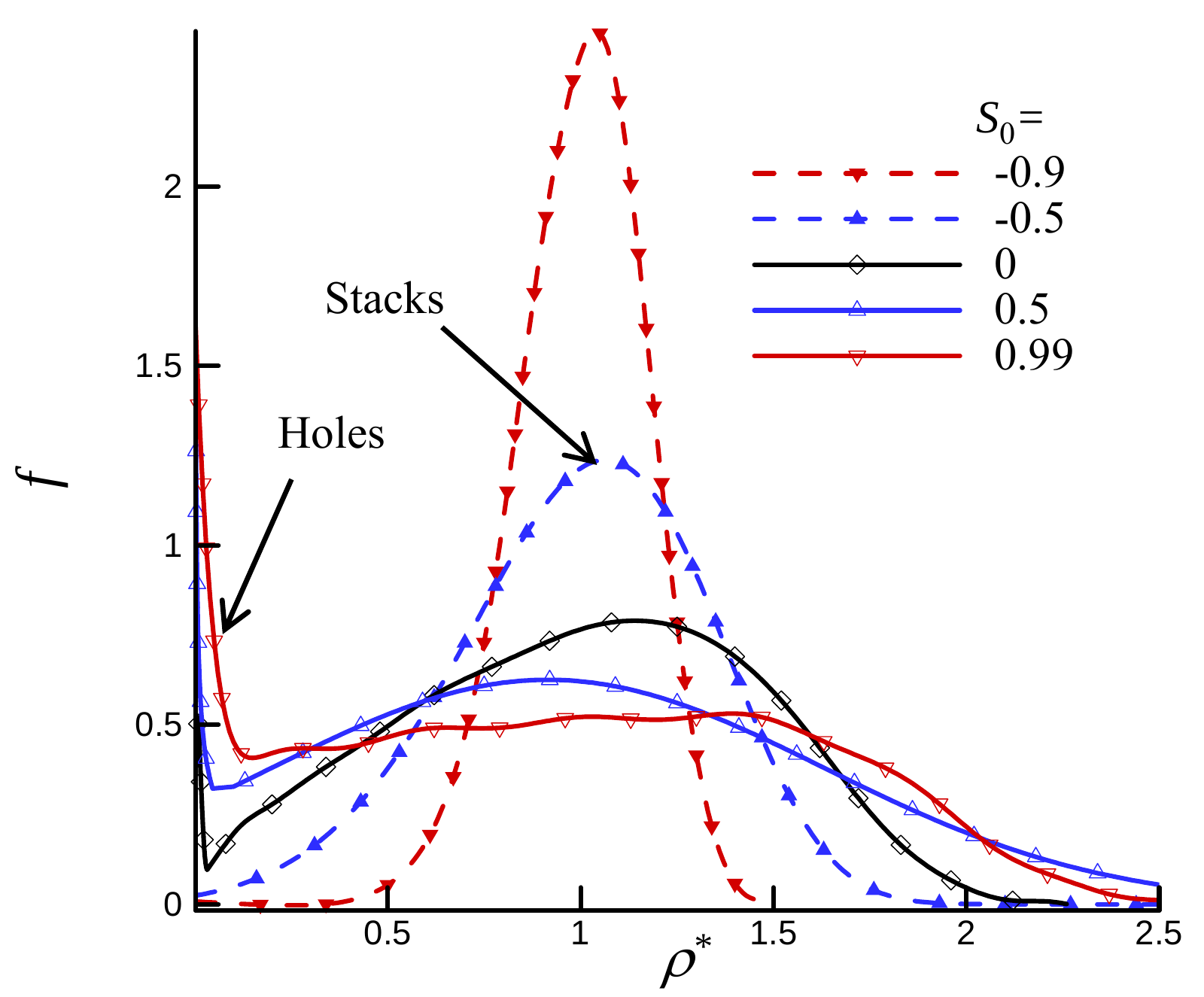}
	\caption{Differential distribution functions of number density, $f$, versus the reduced number density in cells, $\rho^\ast = \rho/\rho_0$ at different preassigned order parameters, $S_0$. The data are presented for the deposition of sticks ($\varepsilon = \infty$) on a line with length $L=2^{15}$. \label{fig:f06}}
\end{figure}

To estimate the inhomogeneities of the number densities for packing of the sticks, the line was divided into $L$ cells; the local number density in each cell was evaluated and the values of the differential distribution function $f(\rho)$ were estimated. For systems with strong ordering along the vertical axis (e.g., for $S_0= -0.9$ in Fig.~\ref{fig:f06})  narrow distributions with maximums located near the value $\rho^\ast = \rho/\rho_0 \approx 1$ were observed. Such distributions correspond mainly to the formation of almost vertical stacks. At $S_0=0$, the distribution function became rather more broad, moreover for particles oriented along the horizontal direction (i.e., $S_0>0$) pronounced peaks located at $\rho^\ast <0.1$ were also observed. These peaks may be attributed to the formation of significant holes between the stacks.

\subsection{Discorectangles}

\begin{figure}[!htbp]
	\centering
	\includegraphics[width=\columnwidth]{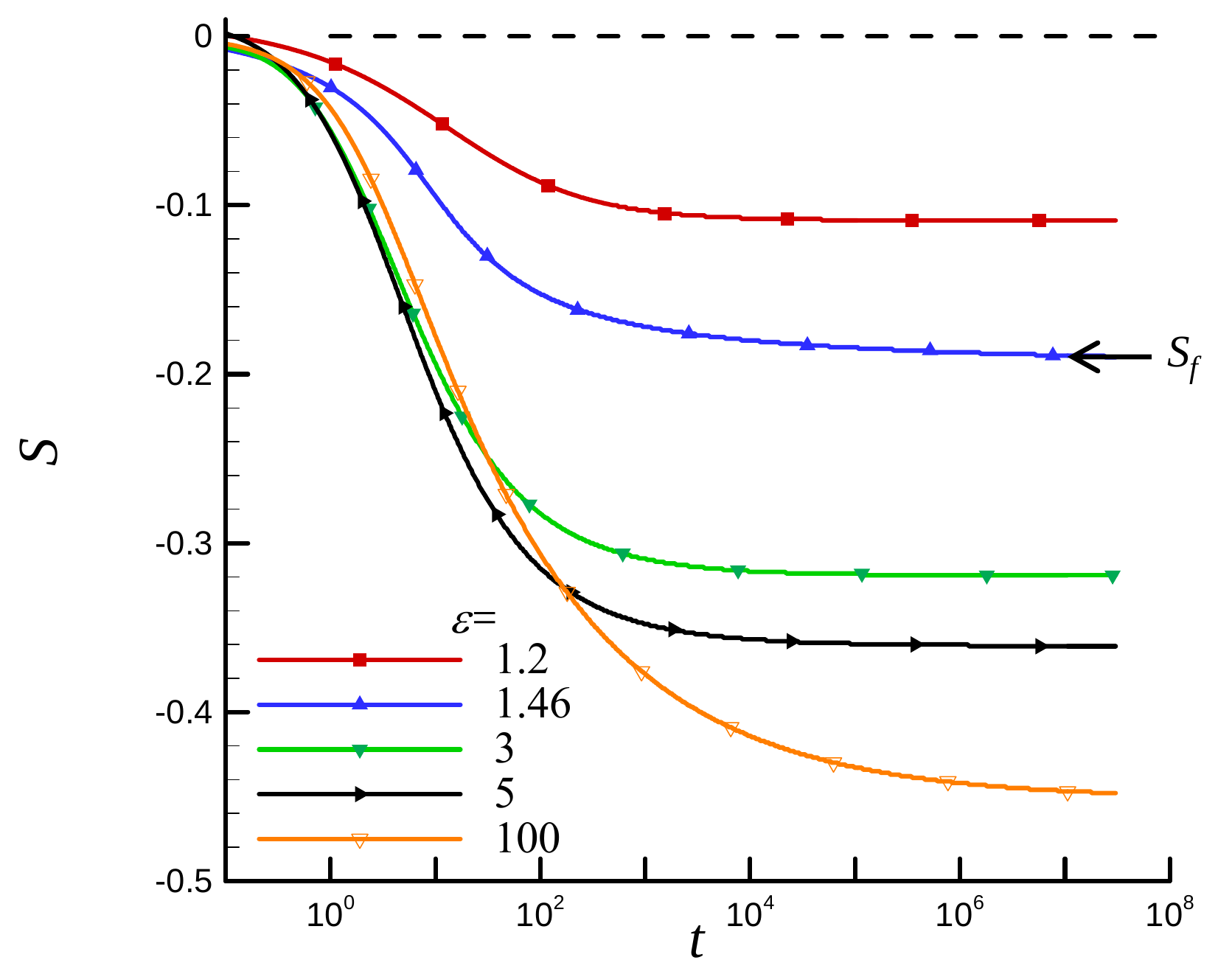}
	\caption{Order parameter, $S$, versus the deposition time, $t$, for disordered RSA packings ($S_0=0$). The data are presented for the deposition of discorectangles with aspect ratio $\varepsilon$ on a line with length $L=2^{15}$. \label{fig:f07}}
\end{figure}

Figure~\ref{fig:f07} shows examples of the order parameter, $S$, versus the deposition time, $t$, for disordered RSA packings ($S_0=0$) at different aspect ratios, $\varepsilon$,  of discorectangles. Similarly to the case of sticks (Fig.~\ref{fig:f03}), the order parameters, $S$, gradually decreased with increasing time, $t$, approaching the value $S_\text{f}$ in the limit of an infinitely large time, $t \to \infty$.

\begin{figure}[!htbp]
	\centering
	\includegraphics[width=\columnwidth]{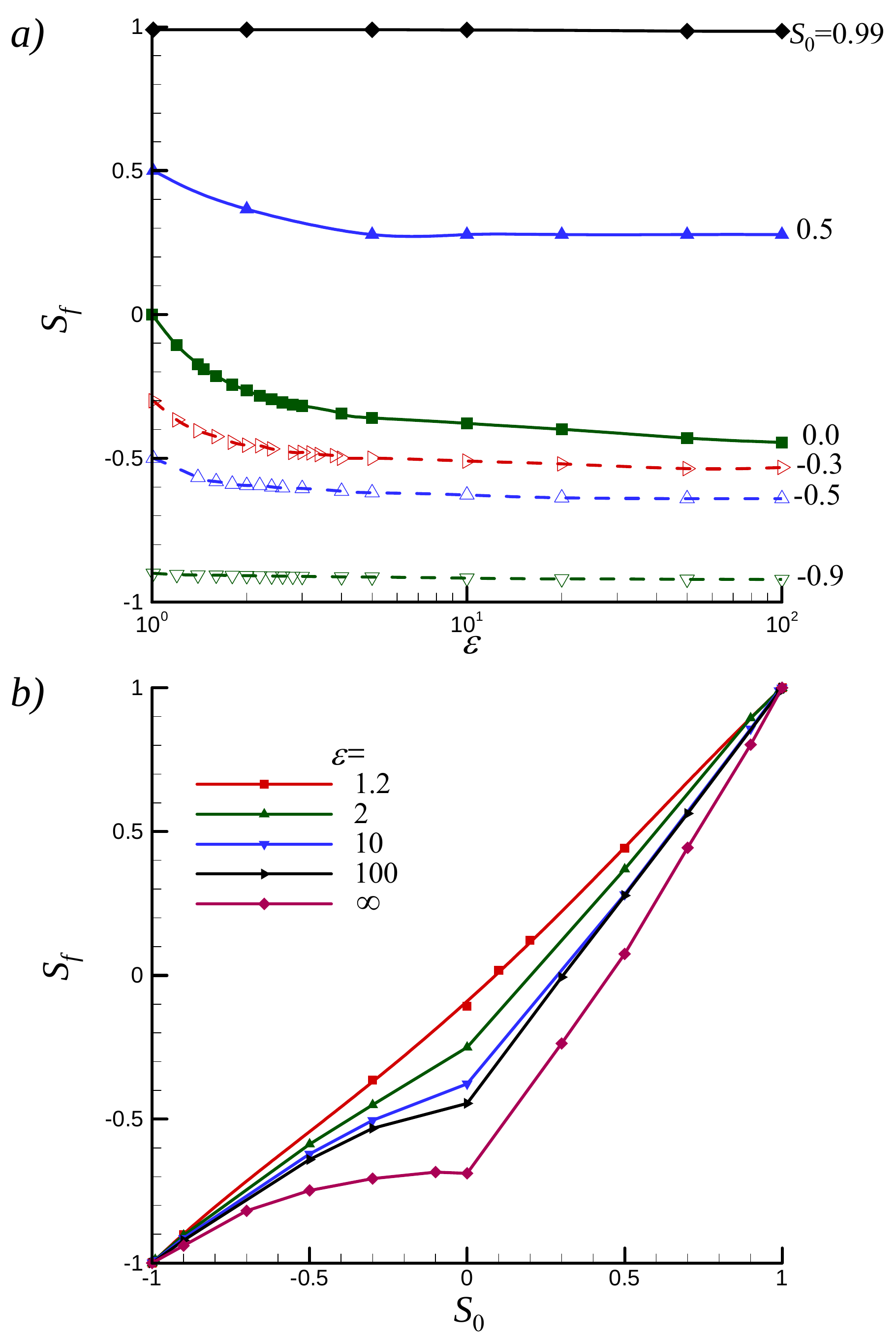}\\
	\caption{Final order parameter in the jamming state, $S_\text{f}$, versus the aspect ratio, $\varepsilon$, at different values of the preassigned order parameter, $S_0$, (a) and $S_\text{f}$ versus $S_0$ at different values of $\varepsilon$. The data are presented for the deposition of discorectangles on a line with length $L=2^{15}$. \label{fig:f08}}
\end{figure}

The changes in order parameter during the deposition can reflect the filtering properties of the RSA deposits. A similar effect was observed for the deposition of sticks (Fig.~\ref{fig:f03}). Similar filtering properties were also observed at other values of $S_0$ and were more significant for elongated particles with large values of $\varepsilon$. For example, the limiting order parameter in the jamming state $S_\text{f}$ ($t \to \infty$) was dependent on the values of the preassigned order parameter $S_0$ and the aspect ratio, $\varepsilon$ (Fig.~\ref{fig:f08}). Note, that in limiting cases of ideal ordering, i.e., at $|S_0|=1$, the values $S_\text{f}$ were unchanged, and in other cases the values of $S_\text{f}$ decreased with increasing $\varepsilon$ (Fig.~\ref{fig:f08}a). For preassigned ordering in the horizontal direction along a line, i.e., at $S_0>0$, the $S_\text{f}(S_0)$ dependencies were almost linear, whereas for ordering in the vertical direction, i.e., at $S_0<0$ these dependencies were non-linear.

\begin{figure}[!htbp]
	\centering
	\includegraphics[width=\columnwidth]{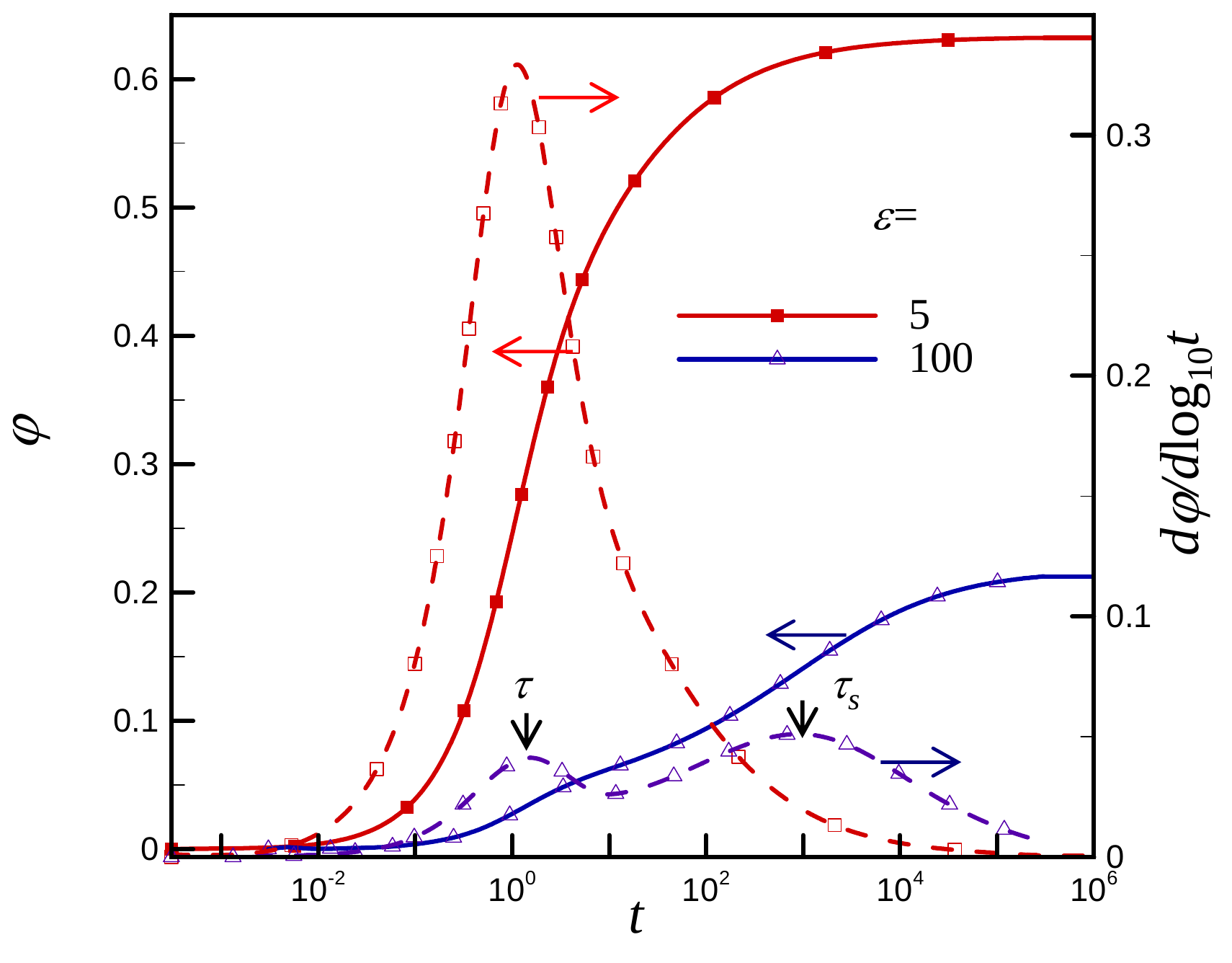}
	\caption{Packing density,  $\varphi$, versus the deposition time, $t$, for disordered RSA packings ($S_0=0$). The time derivatives  $\mathrm{d} \varphi/\mathrm{d}\log_{10} t$ were calculated to estimate the characteristic deposition times $\tau$ and $\tau_s$. The data are presented for the deposition of discorectangles with aspect ratio $\varepsilon$ on a line with length $L=2^{15}$. \label{fig:f09}}
\end{figure}

Figure~\ref{fig:f09} shows examples of the packing density, $\varphi$, versus the deposition time, $t$, for disordered RSA packings ($S_0=0$) at different values of the aspect ratio, $\varepsilon$. The packing density, $\varphi$, gradually increased with increasing time, $t$, approaching the jamming value $\varphi_\text{j}$ at $t \to \infty$. The time derivatives $\mathrm{d} \varphi/\mathrm{d}\log_{10} t$ were also calculated to evaluatethe  inflections at the time dependencies of $\varphi$ (Fig.~\ref{fig:f09}). These inflections were used to estimate the characteristic deposition times, $\tau$. At relatively small values of $\varepsilon$ ($\varepsilon \leqslant 15$) only one inflection point was observed. However, for elongated particles with  $\varepsilon > 20$  two inflection points could be seen (at $\tau$ and $\tau_s$) and this may reflect the development of fast and slow deposition processes (Fig.~\ref{fig:f09}).

\begin{figure}[!htbp]
	\centering
	\includegraphics[width=\columnwidth]{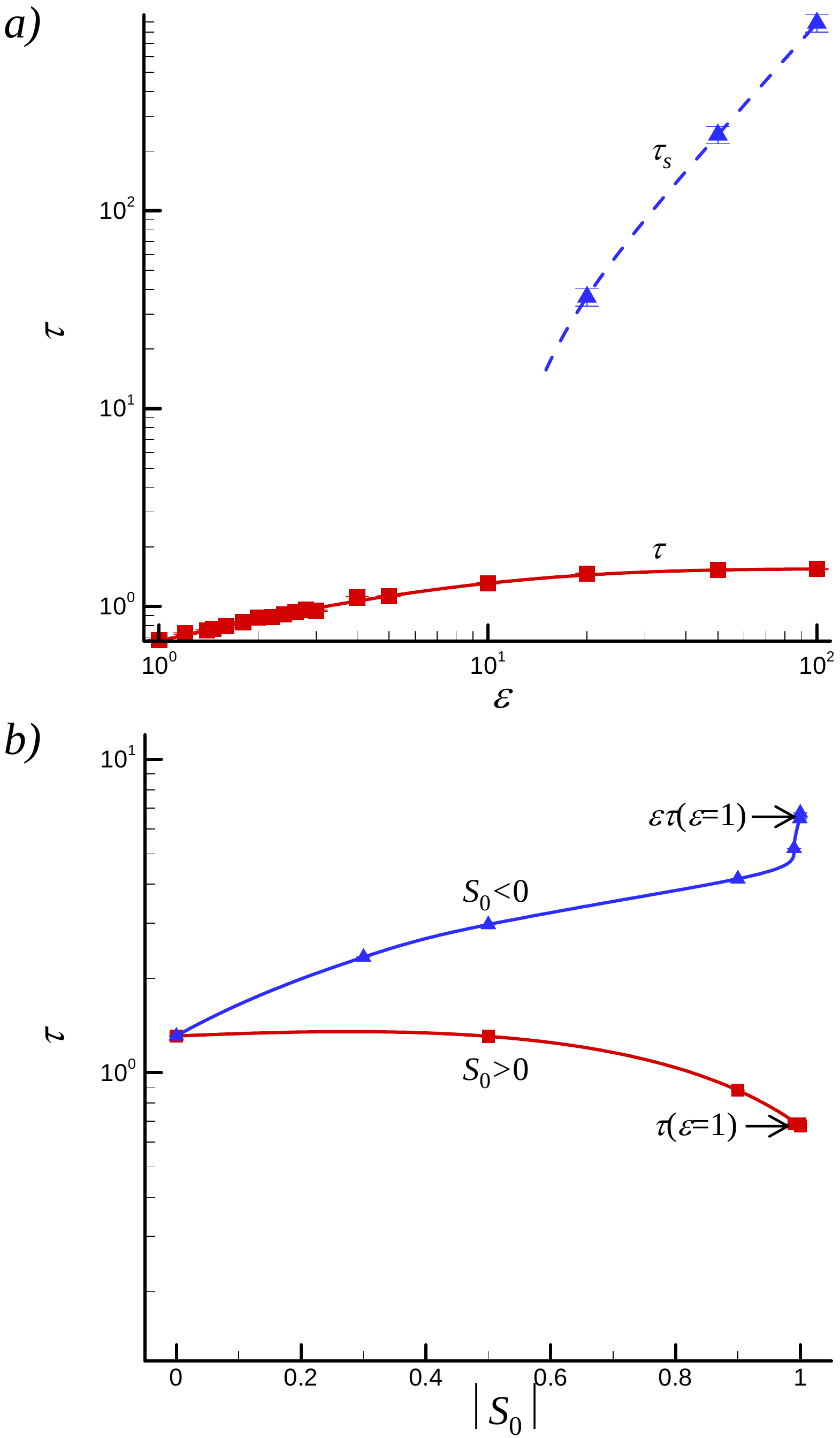}\\
	\caption{Characteristic RSA deposition times $\tau$ and $\tau_s$ versus the aspect ratio of discorectangles, $\varepsilon$, for disordered RSA packings ($S_0=0$) (a), $\tau$ versus $S_0$ at fixed value of $\varepsilon=10$ (b). The data are presented for the deposition of discorectangles with aspect ratios $\varepsilon$ on a line with length $L=2^{15}$. \label{fig:f10}}
\end{figure}

Figure~\ref{fig:f10} demonstrates examples of the behavior of characteristic deposition times at different aspect ratios, $\varepsilon$, and preassigned order parameters, $S_0$. For example, at $S_0=0$ (Fig.~\ref{fig:f10}a), the value of $\tau$ gradually grows as $\varepsilon$ approaches the limit of $\tau \approx 1.6$ at $\varepsilon \to \infty$. The characteristic time of the fast process, $\tau$, may correspond to the transition from loose uncorrelated packing at the initial time to the more dense correlated packing at longer time. For relatively short particles with $\varepsilon \leqslant 15$ only this transition was observed (Fig.~\ref{fig:f10}a). However, for long particles with $\varepsilon \geqslant 20$, the second inflection point at $t=\tau_s$ may  reflect the deposition of particles inside stacks of previously deposited particles. Therefore, the first inflection point is related to standard uncorrelated adsorption. The second inflection point at larger times was only observed at relatively large aspect ratios ($\varepsilon >10$) and it is related to the fact that all possible holes are filled and now adsorption can only happen in the stacks. At this stage, for deposition times in the vicinity of the second inflection point, the voids in stacks can be filled by the particles with some specific orientations, and finally the adsorption is slowing down.

For the fast process, the characteristic time $\tau$ displayed different dependencies on the preassigned order parameter at $S_0<0$ and $S_0>0$. These dependencies for particular value of $\varepsilon=10$ are presented in Fig.~\ref{fig:f10}b. In the limit of $|S_0|=1$, the value of $\tau$ tends towards $\varepsilon\tau(\varepsilon=1)$ and $\tau(\varepsilon=1)$ for the ideal ordering of particles with $S_0=-1$ or $S_0=+1$, respectively.

\begin{figure}[!htbp]
	\centering
	\includegraphics[width=\columnwidth]{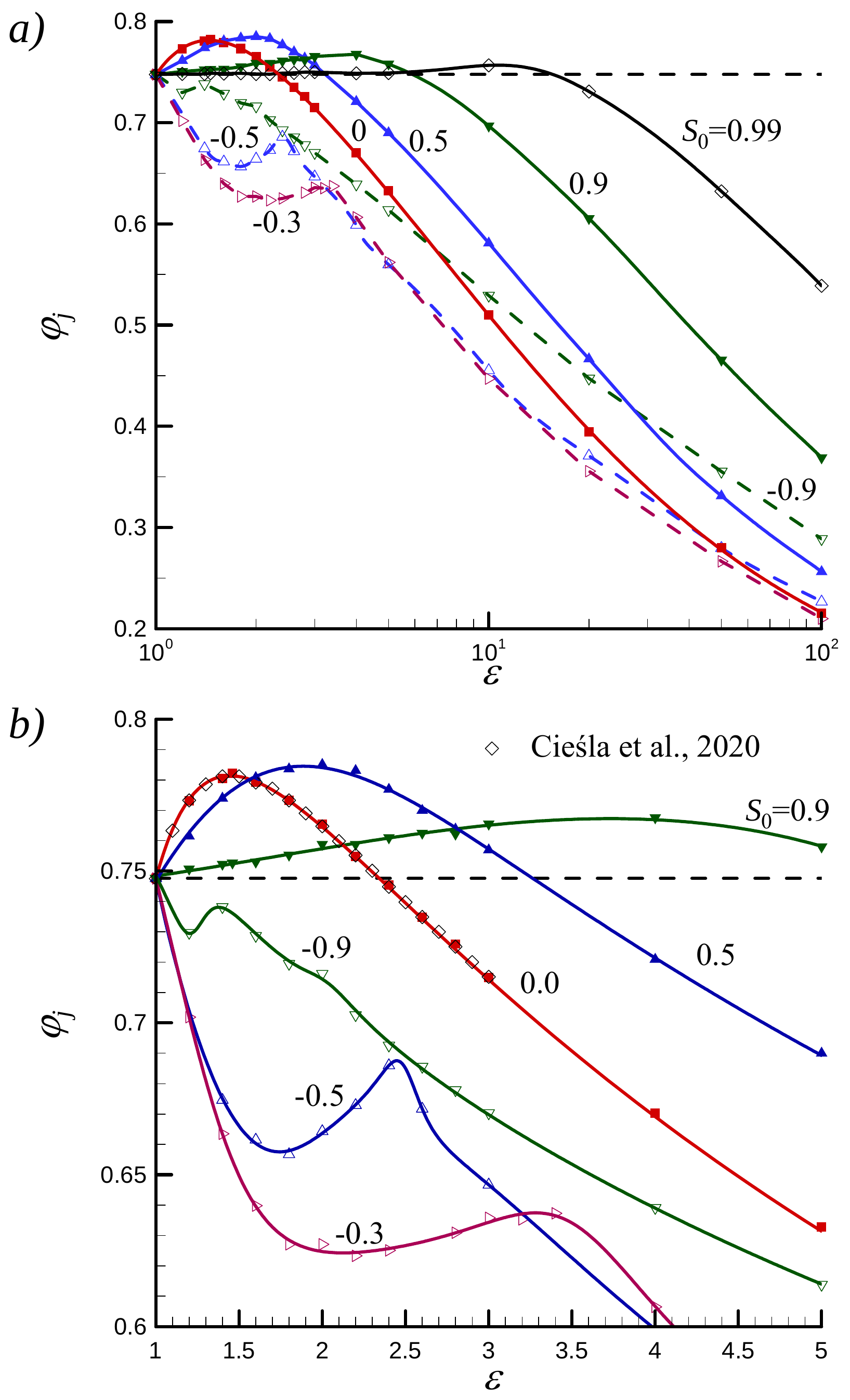}\\
	\caption{Packing density at jamming state, $\varphi_\text{j}$, versus the discorectangle aspect ratio, $\varepsilon$, at different values of the preassigned order parameter, $S_0$, (a) and enlarged portion of the $\varphi_\text{j}(\varepsilon)$ dependencies for $1 \leqslant \varepsilon \leqslant 5$ (b). For $S_0=0$, the maximum $\varphi_\text{j} = 0.7822 \pm 0.004$ was observed at $\varepsilon \approx 1.46$. Dashed horizontal lines correspond to the jamming limit of disks ($\varphi_\text{j}=C_\text{R}=0.7476\dots$ at $\varepsilon \approx 1.0$). The open squares in (b) correspond to the data obtained in~\cite{Ciesla2020}. The data are presented for deposition on a line with length $L=2^{15}$. \label{fig:f11}}
\end{figure}

Figure~\ref{fig:f11} shows the dependences of the packing density at the jamming state, $\varphi_\text{j}$, versus the aspect ratio, $\varepsilon$, at different values of the preassigned order parameters, $S_0$, (a) and enlarged portions of the same dependencies for the interval $1 \leqslant \varepsilon \leqslant 5$ (b). In all cases cusps in the $\varphi_\text{j}(\varepsilon)$ dependencies were observed. For disordered RSA packing (i.e., at $S_0 =0$ a well-defined maximum $\varphi_\text{j}= 0.7822 \pm 0.004$ at $\varepsilon  \approx 1.46$ was observed. Such behavior was in good correspondence with previous data~\cite{Chaikin2006,Ciesla2020}. For example, the obtained values of $\varphi_\text{j}$ for the range of $1 \leqslant \varepsilon \leqslant 3$ were almost the same as those reported in~\cite{Ciesla2020} (Fig.~\ref{fig:f11}b). The initial density increase was explained by relaxing a parameter constraint (appearance of orientational degrees of freedom) in the RSA packing, while  the density decrease at larger values of $\varepsilon$ was explained by the excluded volume effects~\cite{Chaikin2006}. These supplementary degrees of freedom (absent for disks at $\varepsilon = 1$) also significantly affected the algebraic time dependence of the approach of $\varphi_\text{j}$ to jamming~\cite{Baule2017}.

\begin{figure*}[!tb]
	\centering
	\includegraphics[width=\textwidth]{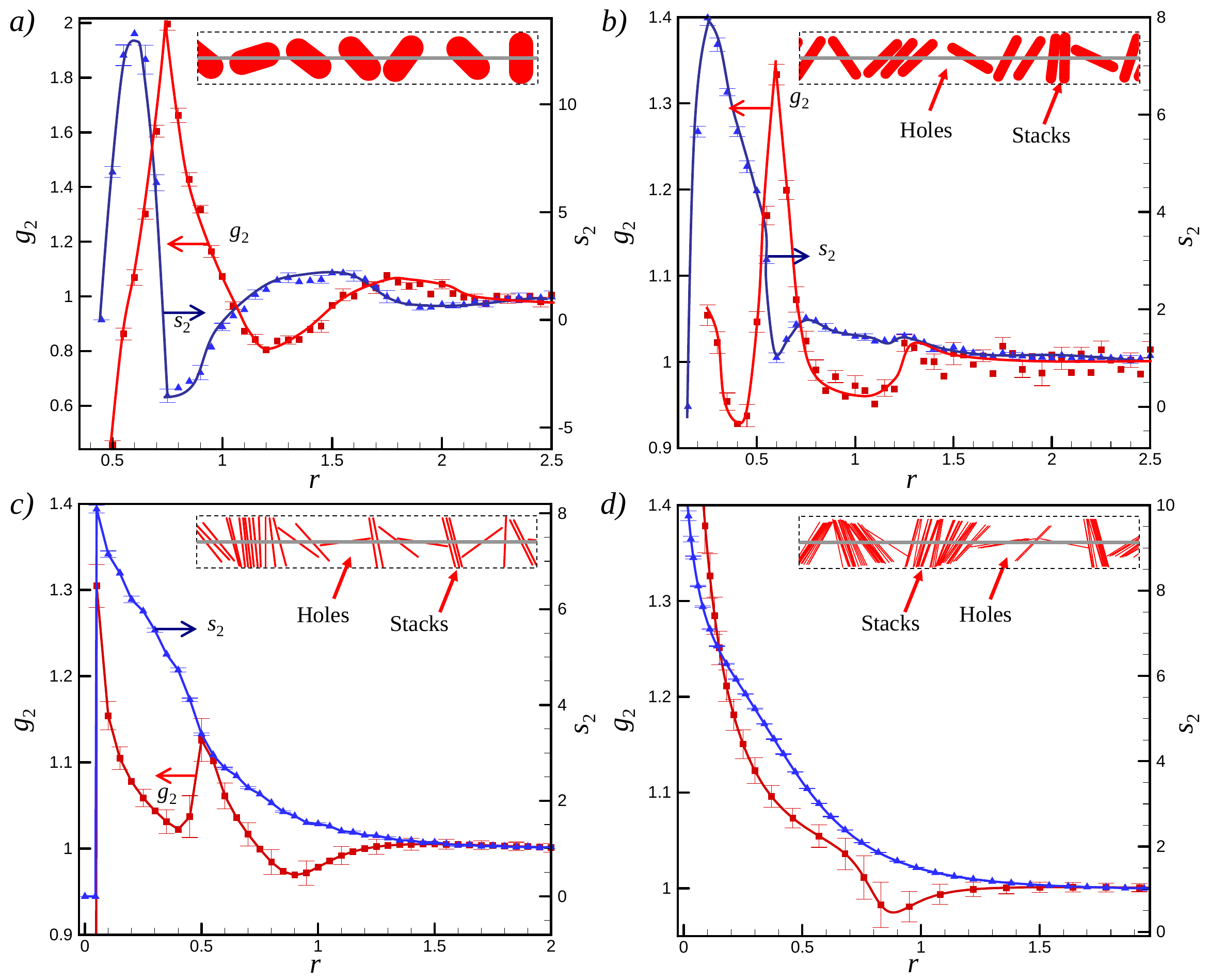}
	\caption{Examples of the density $g_2(r)$ (a) and orientation $s_2(r)$ (b) pair correlation functions with fragments of illustrative RSA disordered packings of discorectangles ($S_0=0.0$) at the jamming concentrations for different aspect ratios $\varepsilon = 2$ (a), $\varepsilon = 5$ (b), $\varepsilon = 20$ (c), and $\varepsilon = 100$ (d). The data are presented for deposition on a line with length $L=2^{15}$.\label{fig:f12}}
\end{figure*}
The preassigned order parameter, $S_0$, noticeably influenced the character of the $\varphi_\text{j}(\varepsilon)$ dependencies and locations of the cusps. At $S_0>0$, the increase in $S_0$ resulted in a shift of the maximum position toward to the larger values of $\varepsilon$. For example, at $S_0=0.5$ the maximum $\varphi_\text{j} \approx 0.785$ was observed at $\varepsilon \approx 1.89$, and at $S_0=0.9$ the maximum $\varphi_\text{j} \approx 0.767$ was observed at $\varepsilon \approx 3.72$. The limit of $S_0 \to 1$ corresponds to the degeneration of the problem to the case with $\varepsilon =1$ (dashed lines in Fig.~\ref{fig:f09}a,b). At $S_0<0$, even more complicated non-monotonic $\varphi_\text{j} (\varepsilon)$ behaviors were observed (Fig.~\ref{fig:f11}b). At small values of $\varepsilon$ (close to $\varepsilon = 1$) noticeable drops in the $\varphi_\text{j}$ values were observed, and then the curves went through their maximums. For elongated particles, at large aspect ratios the excluded volume effects dominated and the density $\varphi_\text{j}$ continuously decreased.

Figure~\ref{fig:f12} presents examples of the density $g_2(r)$ (a) and orientation $s_2(r)$ (b) pair correlation functions at jamming states for different values of the aspect ratio, $\varepsilon$, and a fixed preassigned order parameter, $S_0=0$. At relatively small aspect ratios (Fig.~\ref{fig:f12}a,b) both the functions $g_2(r)$ and $s_2(r)$ exhibited rather complicated and large oscillations at $r \lessapprox 2$, but these became small at $r>3$. These oscillations reflect the size of the correlated regions in the RSA packing of the discorectangles. Note, that the location of the first peak in the density correlation function $g_2(r)$ corresponds to the mean distance between the particles. This should be when the distance is in the order of 1 or $\varepsilon^{-1}$ for ideal horizontal ($S_0=1$) or vertical ($S_0=-1$) RSA packings, respectively. For $\varepsilon = 2$, the location of the first peak of the $g_2(r)$ function was $r \approx 0.75$ and that reflects the noticeable contribution of particles with horizontal orientations, but the location of the first peak of the $s_2(r)$ function was close to $r \approx 0.6$, corresponding  to the closest distance between inclined particles. For $\varepsilon = 5$, the location of the first peak of the $g_2(r)$ function was $r \approx 0.23$ which is very close to $\varepsilon^{-1}=0.2$. It reflects the contribution from particle in parallel stacks clearly visible in fragments of the illustrative RSA packings (Fig.~\ref{fig:f12}). However, the particles with almost horizontal orientations make an important contribution to the second peak located at $r \approx 0.6$. This peak corresponds to the correlation of alternative regions: dense regions filled with parallel stacks and rarefied regions (holes) containing particles with almost horizontal orientations. For this particular case of $\varepsilon = 5$ the behavior of the $g_2(r)$ and $s_2(r)$ functions was approximately anti-bat. At larger values of $\varepsilon$ the first peaks in $g_2(r)$ and $s_2(r)$ became located at $r \approx \varepsilon^{-1}$ reflecting the dominant contribution from particles in parallel stacks (see corresponding fragments of the illustrative RSA packings in Fig.~\ref{fig:f12}c,d). The contribution from holes (correlation of alternative regions) became less important with increased values of $\varepsilon$.

\section{Conclusion\label{sec:conclusion}}

Numerical studies of two-dimensional RSA deposition of infinitely thin particles (sticks) and discorectangles on a one-dimensional line were performed. The packing kinetics and properties of the packs were significantly influenced by the values of the preassigned order parameter, $S_0$, and the aspect ratio, $\varepsilon$. The deposition was governed by the formation of rarefied holes (containing particles oriented along the line) surrounded by comparatively dense stacks (filled by almost parallel particles oriented in the vertical direction). This resulted in significant deviation of the actual order parameter $S$ in the deposit and of the preassigned order parameter $S_0$. In fact, the unsaturated packing acted as a filter for the adsorption of particles with appropriated orientations. For the RSA packing of discorectangles, the filtering properties of the RSA deposits were more significant at relatively large aspect ratios. For elongated particles with $\varepsilon \geqslant 20$, the development of fast and slow deposition processes could be  observed. However, the introduction of preferential ordering also influenced the behavior of the cusps observed in the $\varphi_\text{j}(\varepsilon)$ dependencies in the interval $1 \leqslant \varepsilon \leqslant5$. The observed effects can be explained by the impact of the partial ordering on the competition between the orientational degrees of freedom of each particle and by excluded volume effects~\cite{Chaikin2006}.

\begin{acknowledgments}
We acknowledge funding from the National Academy of Sciences of Ukraine, Project Nos.~0117U004046 and 0120U100226 (7/9/3-f-4-1230-2020) (N.I.L., N.V.V.), and the Russian Foundation for Basic Research, Project No.~18-07-00343 (Yu.Yu.T.).
\end{acknowledgments}

\bibliography{Paris2020}

\end{document}